\documentclass[sn-nature]{sn-jnl}% Style for submissions to Nature Portfolio journals
\usepackage{graphicx}%
\usepackage{multirow}%
\usepackage{amsmath,amssymb,amsfonts}%
\usepackage{amsthm}%
\usepackage{mathrsfs}%
\usepackage[title]{appendix}%
\usepackage{xcolor}%
\usepackage{textcomp}%
\usepackage{manyfoot}%
\usepackage{booktabs}%
\usepackage{algorithm}%
\usepackage{algorithmicx}%
\usepackage{algpseudocode}%
\usepackage{listings}%
\theoremstyle{thmstyleone}%
\theoremstyle{thmstyletwo}%

\theoremstyle{thmstylethree}%

\begin{document}

\title[Article Title]{LLM-based Multi-Agent Copilot for Quantum Sensor}

%%=============================================================%%
%% GivenName	-> \fnm{Joergen W.}
%% Particle	-> \spfx{van der} -> surname prefix
%% FamilyName	-> \sur{Ploeg}
%% Suffix	-> \sfx{IV}
%% \author*[1,2]{\fnm{Joergen W.} \spfx{van der} \sur{Ploeg} 
%%  \sfx{IV}}\email{iauthor@gmail.com}
%%=============================================================%%
\author[1,2]{\fnm{Rong} \sur{Sha}}
\equalcont{These authors contributed equally to this work.}
%\email{shar20@nudt.edu.cn}
\author[1]{\fnm{Binglin} \sur{Wang}}
\equalcont{These authors contributed equally to this work.}
%\email{wangbinglin14@nudt.edu.cn}
\author*[1,2]{\fnm{Jun} \sur{Yang}}\email{jyang@nudt.edu.cn}
\equalcont{These authors contributed equally to this work.}

\author[1,2]{\fnm{Xiaoxiao} \sur{Ma}}
\equalcont{These authors contributed equally to this work.}
%\email{maxiaoxiao@zju.edu.cn}
\author[1]{\fnm{Chengkun} \sur{Wu}}
%\email{chengkun$\_$wu@nudt.edu.cn}
\author[1]{\fnm{Liang} \sur{Yan}}
%\email{yanliang@nudt.edu.cn}
\author[1,2]{\fnm{Chao} \sur{Zhou}}

\author[1,2]{\fnm{Jixun} \sur{Liu}}
%\email{liujixun@buaa.edu.cn}
\author[1,2]{\fnm{Guochao} \sur{Wang}}

\author[1,2]{\fnm{Shuhua} \sur{Yan}}
%\email{yanshuhua996@nudt.edu.cn}
\author*[1,2]{\fnm{Lingxiao} \sur{Zhu}}\email{lxz334@nudt.edu.cn}

\affil*[1]{\orgname{National University of Defense Technology}, \orgaddress{\city{Changsha}, \postcode{410073} \country{China}}}

\affil[2]{\orgdiv{National Key Laboratory of Equipment State Sensing and Smart Support}, \orgname{National University of Defense Technology}, \orgaddress{\city{Changsha}, \postcode{410073} \country{China}}}

%%==================================%%
%% Sample for unstructured abstract %%
%%==================================%%
%
\abstract{
Large language models (LLM) exhibit broad utility but face limitations in quantum sensor development, stemming from interdisciplinary knowledge barriers and involving complex optimization processes. Here we present QCopilot, an LLM-based multi-agent framework integrating external knowledge access, active learning, and uncertainty quantification for quantum sensor design and diagnosis. Comprising commercial LLMs with few-shot prompt engineering and vector knowledge base, QCopilot employs specialized agents to adaptively select optimization methods, automate modeling analysis, and independently perform problem diagnosis. Applying QCopilot to atom cooling experiments, we generated 10${}^{\rm{8}}$ sub-$\rm{\mu}$K atoms without any human intervention within a few hours, representing $\sim$100$\times$ speedup over manual experimentation. Notably, by continuously accumulating prior knowledge and enabling dynamic modeling, QCopilot can autonomously identify anomalous parameters in multi-parameter experimental settings. Our work reduces barriers to large-scale quantum sensor deployment and readily extends to other quantum information systems.}

\maketitle

\section{Introduction}\label{sec1}
%, quantum computing\cite{RevModPhys.82.2313}, quantum simulation\cite{Gruneisen2021} 
Atom cooling, trapping and quantum manipulation has been one of the leading platforms in quantum sensing in the last decades\cite{wiemanAtomCoolingTrapping1999,beckerSpaceborneBoseEinstein2018,schreckLaserCoolingQuantum2021,Yuan_2023,huangEntanglementenhancedQuantumMetrology2024,SalducciSA2024,pandaMeasuringGravitationalAttraction2024,JunPRL2024}. Cold-atoms based quantum sensors are increasingly transitioning from controlled laboratory environments to real-world deployments\cite{barrettDualMatterwaveInertial2016,bidelAbsoluteMarineGravimetry2018,Bongs:2019exm,XuejianSA2019,strayQuantumSensingGravity2022,darmagnacdecastanetAtomInterferometryArbitrary2024,dengColdAtomMicrowave2024,zhaiAirborneAbsoluteGravity2025,26vs-4t7p2025,LiNSR2025}. However, such systems—integrating laser subsystems, optical components, electronic control modules, and firmware—exhibit high complexity in their operational dynamics\cite{DegenRevModPhys2017}. Furthermore, cold atoms, as the core sensing medium, display extreme susceptibility to thermal perturbations, mechanical vibrations, and electromagnetic noise\cite{geraciSensitivityAtomInterferometry2016}. This confluence of complexity and environmental sensitivity elevates deployment costs and degrades operational robustness. Developing cold-atoms based devices capable of self-optimization and self-diagnosis during prolonged field operation thus remains a critical bottleneck. Existing strategies fall into three broad categories: control-based methods (e.g., PID control, robust control\cite{saywellEnhancingSensitivityAtominterferometric2023a}), black-box optimization techniques (e.g., Bayesian optimization\cite{2016NatSR...625890W,nakamuraNonstandardTrajectoriesFound2019,Barker_2020,maAcceleratedBayesianOptimization2025},  genetic algorithms\cite{weidnerAtomInterferometryUsing2017}, deep neural networks\cite{2018NatCo...9.4360T}), and reinforcement learning (RL)-based agents\cite{AlexeyPNAS2018,2023AdPho...5a6005C,chihReinforcementLearningRotation2024,2024NatCo..15.8532R,Liang24}.  Control-based approaches offer strong interpretability via mathematical models but falter when models are inaccurate or systems exhibit nonlinearity—a common scenario in real-world experiments. Black-box methods, while model-agnostic, suffer from limited generalization due to data quality dependencies and domain specificity\cite{wangScientificDiscoveryAge2023}.  RL-based agents enable autonomous learning but rely on high-quality offline datasets and human-designed reward functions, stumbling over the Polanyi Paradox—the gap between tacit knowledge and explicit articulation. A unifying limitation across these methods is their inability to bridge knowledge barriers, hindering scalable utility and cumulative knowledge growth, and creating a critical trade-off between generalization and domain adaptability.

Large language model (LLM) offers a transformative solution to these challenges.  Pre-trained on vast datasets, LLM internalizes cross-disciplinary knowledge,  exhibits strong few-shot and zero-shot generalization, and excels in reasoning,  planning, and contextual understanding\cite{singh_progprompt_2023}. Unlike RL agents, LLM-based agents operate via text instructions, prompt templates, and in-context learning, enabling flexible,  human-like interaction\cite{boiko_autonomous_2023,gaoEmpoweringBiomedicalDiscovery2024}. When integrated into multi-agent systems, LLMs foster collective intelligence, as demonstrated in software development\cite{Dong2024}, multi-robot planning\cite{Chen2024}, self-driving experiment optimization\cite{sandersBiologicalResearchSelfdriving2023,m_bran_augmenting_2024,luoLargeLanguageModels2024,ruanAutomaticEndtoendChemical2024,jablonkaLeveragingLargeLanguage2024,wu_leveraging_2024}, and scientific discovery\cite{tshitoyanUnsupervisedWordEmbeddings2019,melkoLanguageModelsQuantum2024,ghafarollahiSciAgentsAutomatingScientific2024}. However, existing LLM-based frameworks lack integration across real-time monitoring, experimental optimization and fault diagnosis—critical for quantum experiments, which are inherently uncertain.

\begin{figure}[h]
\centering
\includegraphics[width=0.95\textwidth]{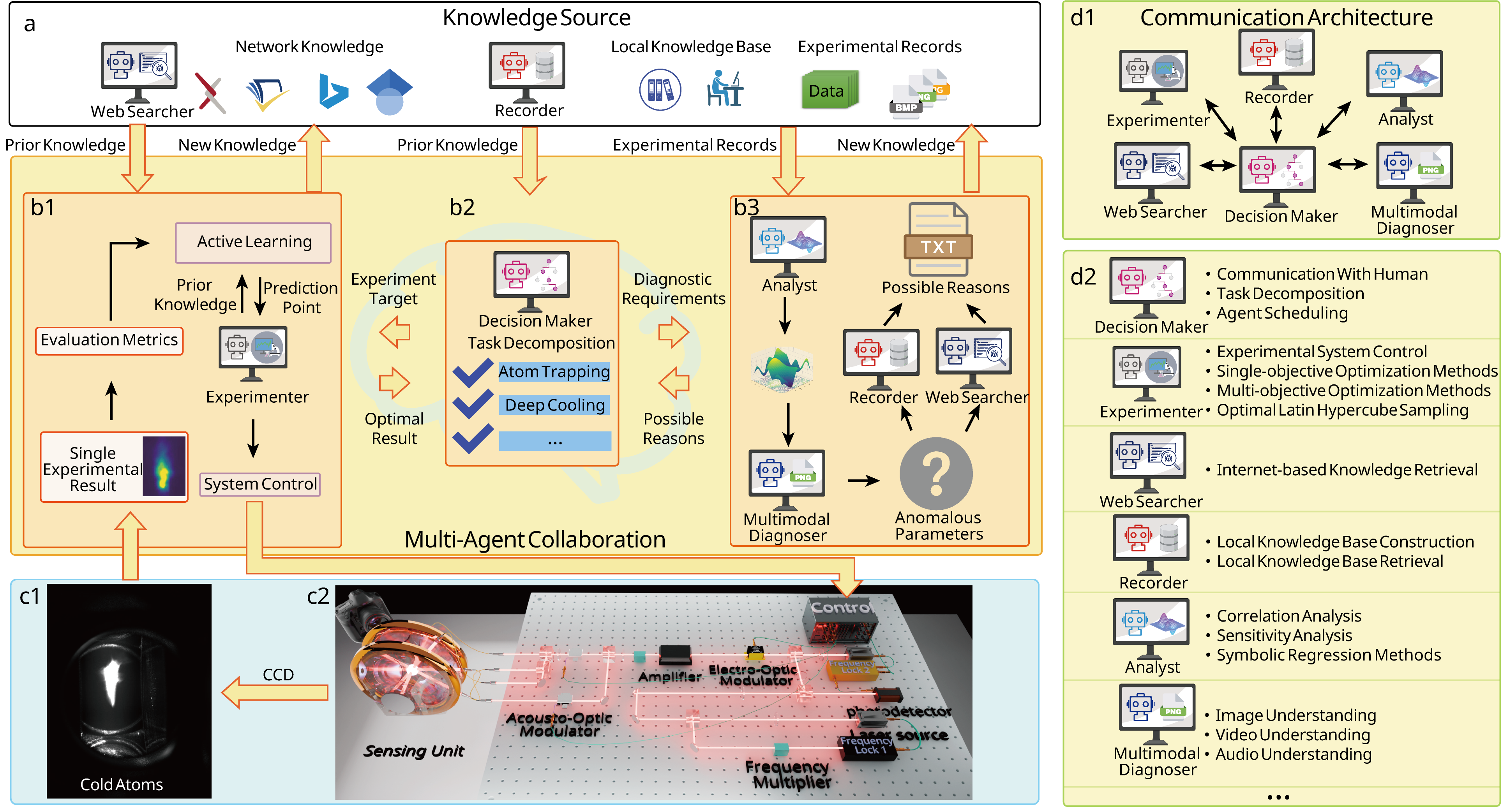}
\caption{\textbf{Architecture of QCopilot}$\quad$ \textbf{a}, The system employs the Web Searcher and Recorder to construct a centralized knowledge source, integrating both historical experimental data and external information. \textbf{b2}, During task decomposition, the Decision Maker utilizes prior knowledge from this centralized source to decompose complex tasks and determine optimal agent invocation workflows. \textbf{b1}, Experimental optimization workflow: The Experimenter optimizes experimental parameters through active learning techniques, driving autonomous system tuning. \textbf{b3}, Fault diagnosis workflow: The Analyst models system behavior, the Multimodal Diagnoser performs image-based anomaly analysis, and the Recorder (with the Web Searcher) retrieves potential root causes to enable targeted troubleshooting. \textbf{c1}, Cold-atom fluorescence image captured by charge coupled device (CCD). \textbf{c2}, Schematic of the experimental setup, illustrating key components of the cold-atom platform. \textbf{d1}, The framework adopts a centralized agent communication architecture to coordinate inter-agent interactions. \textbf{d2}, Inventory of existing agents, highlighting the framework's modular design that supports flexible integration of additional agents as needed.}\label{fig1}
\end{figure}

To address these gaps, we introduce QCopilot, a centralized multi-agent framework (Fig. \ref{fig1}\textbf{d1}) that combines LLM (e.g., Qwen3), vector knowledge bases, active learning, and uncertainty quantification. This framework enables bidirectional functionality: forward optimization of experimental systems and reverse diagnosis of anomalies. As shown in Fig. \ref{fig1}\textbf{d2}, QCopilot orchestrates specialized agents—Decision Maker, Experimenter, Analyst, Multimodal Diagnoser, Web Searcher, and Recorder—to decompose tasks, automate optimization, quantify uncertainties, and diagnose faults. By synergistically integrating its core components, the framework effectively breaks down knowledge barriers, leverages natural language-based prior knowledge to enhance decision-making, and continuously refines and accumulates knowledge, as illustrated in Fig. \ref{fig1}\textbf{a} and \textbf{b1-b3}. During the experimental optimization stage that shown in Fig. \ref{fig1}\textbf{b1}, Experimenter directly interacts with a cold atoms experimental platform (Fig. \ref{fig1}\textbf{c1}-\textbf{c2}) and employs active learning techniques to optimize experimental parameters. Upon completion of the optimization, updated experimental records and associated knowledge are integrated into the knowledge source. Experimental validation in atomic cooling demonstrates the generation of 10$^{\rm{8}}$ sub-$\rm{\mu K}$ atoms within several hours, achieving a $\sim$100-fold speedup over manual experimentation. In the problem diagnosis stage that displayed in Fig. \ref{fig1}\textbf{b3}, Analyst processes experimental records, which are transformed into visual data and forwarded to the Multimodal Diagnoser to identify anomalous parameters. Working in collaboration with the Recorder and Web Searcher, the system infers potential root causes, addresses the issues, and subsequently updates the knowledge base with newly acquired insights. This framework is capable of autonomously detecting anomalous parameters in multi-parameter experimental environments. As a result, QCopilot accelerates the automated preparation of cold atoms workflows, enhances generalization to environmental variations, and paves the way for the development of cost-effective and reliable cold-atoms based quantum sensors in both academic and industrial settings.

\section{Results}\label{sec2}
In cold-atoms based quantum sensors, dense cold-atoms source at sub-$\rm{\mu K}$ temperatures is typically required for improving sensing precision and sensitivity. To achieve this, the experimental process is divided into two stages: capturing and cooling the atomic ensemble to the sub-mK level using a magneto-optical trap (MOT), followed by further reducing the temperature to the $\rm{\mu K}$ level through sub-Doppler cooling. As illustrated in Fig. \ref{fig1}\textbf{c}, our experimental platform combines commercially available lasers with self-developed control systems. First, we stabilize the laser frequency using modulation transfer spectroscopy and optical phase-locked loop. Then, three pairs of circularly polarized orthogonal lasers, along with a gradient magnetic field, form a MOT for cooling and trapping rubidium atoms within the vacuum chamber. In our system, polarization gradient cooling (PGC) serves as the primary mechanism for sub-Doppler cooling. The advantage of this approach is its simplicity, as it does not require any additional optical or electronic components. After loading the MOT, the magnetic field is turned off after 1s, and the laser frequency and power are adjusted accordingly. To determine the total number of atoms in the MOT, we use pixel integration from charge coupled device (CCD) images, as shown in Fig. \ref{fig2}\textbf{b}. This step aims to acquire a dense atomic cloud. During the PGC stage, we employ the molasses diffusion method to measure the temperature and atom number. Representative images are shown in Fig. \ref{fig2}\textbf{e}, while the fitted results of the atomic cloud diffusion radius are presented below. Our objective is to obtain a cold atomic ensemble with plenty of atoms.

The whole experiments are subject to significant uncertainties. Controllable variables, such as optical parameters, dictate the upper limit of optimization during the tuning process, while uncontrollable variables, such as environmental factors, restrict the system's generalization capabilities. Given the high-dimensional and sparse nature of the parameters, combined with the substantial experimental costs, researchers must carefully balance cost and performance. This often renders parameter selection heavily reliant on prior experimental experience. Furthermore, manual optimization is limited in its ability to simultaneously consider multiple objectives, significantly increasing the complexity of the process. The intricate experimental procedure and time-consuming individual sub-experiments optimizations limit the total number of trials that can be performed. The quality of the atomic cloud produced in the cooling stage directly influences key outcomes such as the signal-to-noise ratio of interference signals. 

\subsection{Adaptive optimization}

\begin{figure}[h]
\centering
\includegraphics[width=0.95\textwidth]{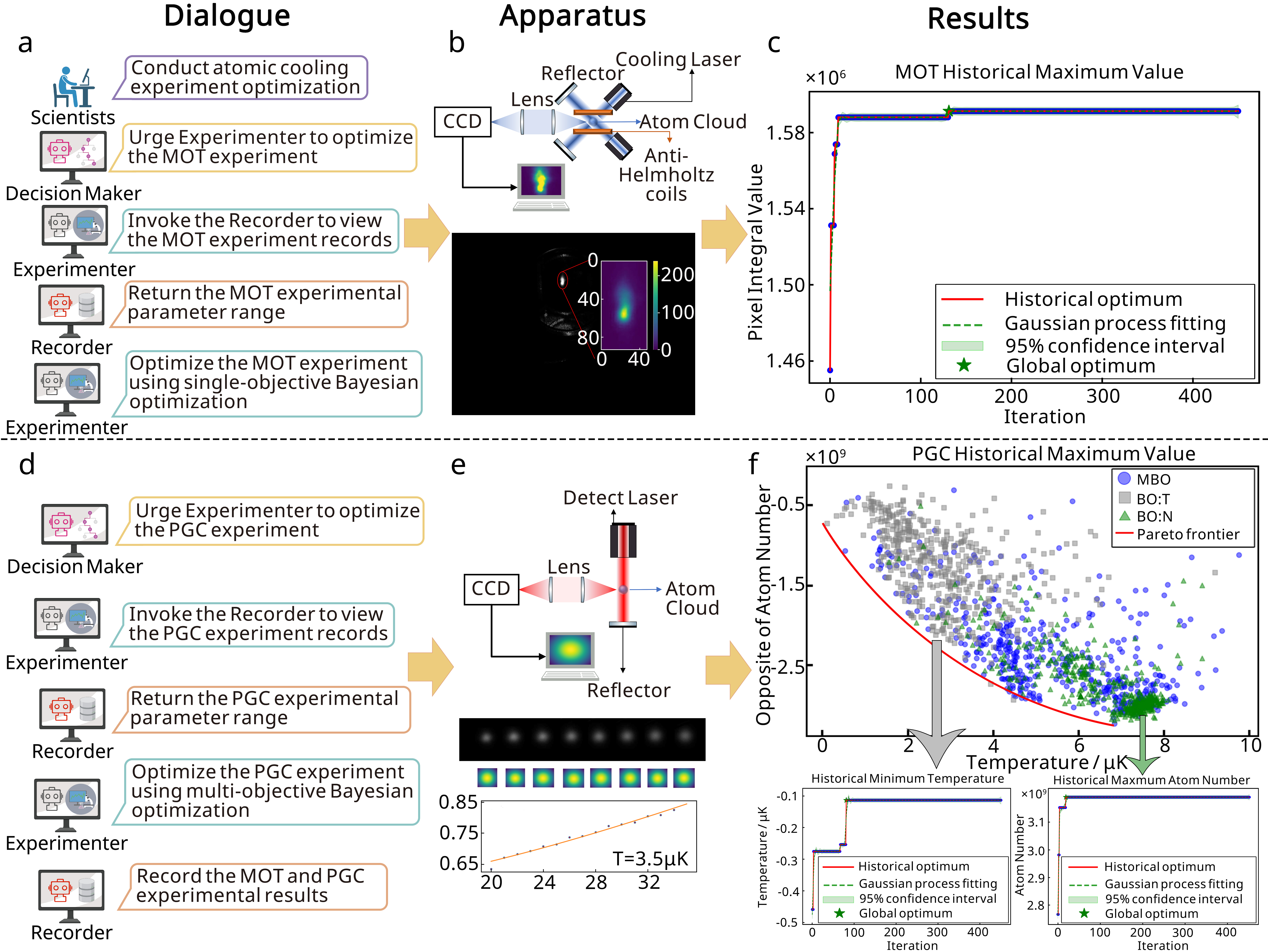}
\caption{\textbf{Optimization of atomic cooling experiments via QCopilot}$\quad$\textbf{a}, Workflow for optimizing the magneto-optical trap (MOT) via the QCopilot framework. \textbf{d}, Corresponding workflow for polarization gradient cooling (PGC) optimization. \textbf{b}, Experimental procedure for quantifying MOT atom number using CCD image pixel integration. \textbf{c}, Evolution of Bayesian optimization (BO) performance in MOT optimization, tracking the historical maximum pixel integral (proportional to atom number). Single-objective BO, targeting five key parameters, converges to optimal values within $\sim$100 iterations. \textbf{e}, Protocol for measuring atomic temperature via the molasses diffusion method. \textbf{f}, Multi-objective optimization results for PGC: main panel shows Pareto frontier (red line) identified by multi-objective Bayesian optimization (MBO, blue points) across 500 iterations, targeting six parameters. Insets display historical optimization trends for minimizing temperature (gray squares) and maximizing atom number (green triangles) via single-objective BO. MBO efficiently balances both objectives, yielding sub-µK atoms with high density, whereas single-objective approaches prioritize one metric at the expense of the other.}\label{fig2}
\end{figure}

For experimental optimization, we provide the Decision Maker with instructions regarding the list of sub-experiments, their order, and specific requirements. The Decision Maker calls the Experimenter sequentially according to the sub-experiment order. Based on the requirements of different sub-experiments, the Experimenter automatically selects appropriate optimization methods and sets experimental parameters using retrieval-augmented generation to retrieve hardware parameter reports from a knowledge base, without requiring prior physical knowledge. From the dialogue history as shown in Fig. \ref{fig2}\textbf{a}, the Experimenter employs Bayesian optimization (BO) based on batch Log Expected Improvement to efficiently search the high-dimensional parameter space for the MOT experiment. To enhance robustness, each parameter setting during optimization was tested three times, with the average value recorded. Fig. \ref{fig2}\textbf{c} shows the historical optimal values across iterations. The ordinate is obtained by performing pixel integration on the CCD image, and the pixel integration value is proportional to the number of atoms. Optimization converged to the optimum in approximately 100 iterations. 

Having determined the best parameters for the MOT, the Decision Maker invokes the Experimenter to fix these parameters for subsequent experiments. For the PGC optimization, the Experimenter uses multi-objective Bayesian optimization (MBO) based on parallel Expected Hypervolume Improvement to identify the Pareto frontier, as displayed in Fig. \ref{fig2}\textbf{d}. Since interferometric measurements generally require cold atomic ensembles below 10 $\rm{\mu K}$, only data points representing temperatures below this threshold are depicted in Fig. \ref{fig2}\textbf{f}. Each comparison experiment consisted of 500 setups, yielding acceptable results (below 10 $\rm{\mu K}$) as follows: 477 Blue circles represent experimental points guided by MBO, while the red line indicate the Pareto frontier deduced by MBO. 463 Gray squares correspond to results achieved when the Decision Maker optimized solely for temperature using single-objective BO. 489 Green triangles depict results when the Decision Maker optimized solely for atom number. Notably, BO optimizes for the maximum, so the temperature was multiplied by -1 to align with the optimization goal. Conversely, in MBO, all objectives were minimized, necessitating multiplication of the atom number by -1. As illustrated in Fig. \ref{fig2}\textbf{f}, points near the origin, which satisfy both low temperature and high atom number, are concentrated due to MBO. Optimization focuses solely on temperature yielded points in regions of low temperature but insufficient atom density, whereas optimization focuses on atom number produced points with high atom number but relatively higher temperature.  MBO targeting six parameters across 500 iterations efficiently identifies the Pareto frontier, which demonstrates the superior performance of MBO in meeting experimental requirements. All three methods demonstrated efficient searching, effectively balancing experimental cost and system performance.

\subsection{Autonomous analysis}

\begin{figure}[h]
\centering
\includegraphics[width=0.95\textwidth]{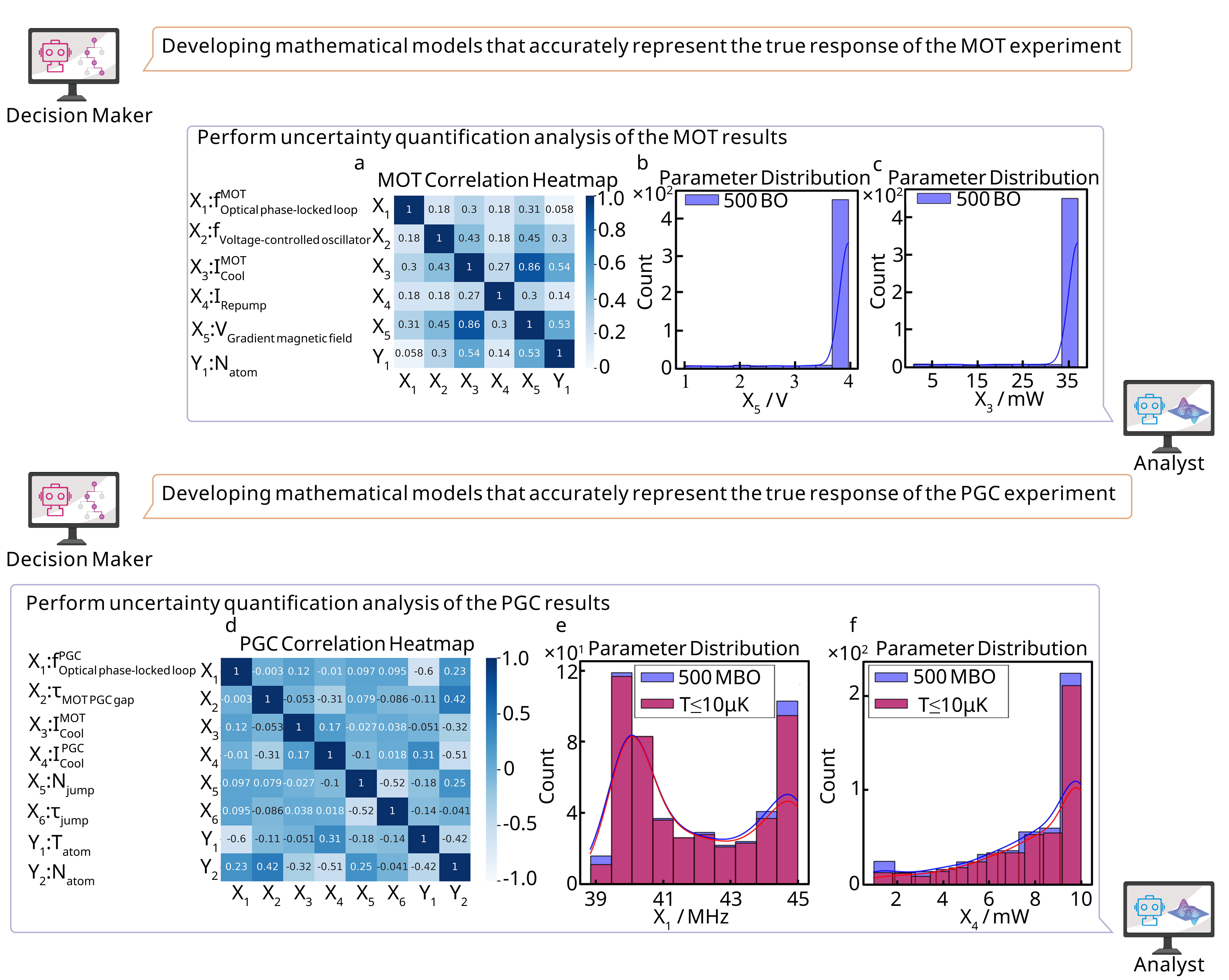}
\caption{\textbf{Uncertainty quantification of the experimental system response}$\quad$The Analyst performed uncertainty quantification on 500 sets of magneto-optical trap (MOT) experimental results generated via Bayesian optimization. \textbf{a}, Correlation matrix quantifying the linear relationships between variables. \textbf{b},\textbf{c}, Distributions of key parameters: voltage of the gradient magnetic field (\textbf{b}) and cooling light intensity (\textbf{c}). Subsequent analysis was conducted on 500 sets of polarization gradient cooling (PGC) results obtained from multi-objective Bayesian optimization. \textbf{d}, Correlation matrix characterizing parameter interactions in the PGC experiment. \textbf{e},\textbf{f}, Distributions of the optical phase-locked loop reference frequency during PGC (\textbf{e}) and the cooling light intensity at the PGC endpoint (\textbf{f}), with probability density curves overlaid to highlight parameter ranges critical for sub-µK cooling performance.}\label{fig3}
\end{figure}

As illustrated in Figs. \ref{fig3}\textbf{a-c}, we perform  correlation analysis, a typical method of uncertainty quantification, to analyze the correlation of parameters in the optimized experimental data from the MOT. Fig. \ref{fig3}\textbf{a} indicates that the number of atoms is strongly correlated with both the gradient magnetic field voltage $\rm{X}_5$ and the cooling light intensity $\mathrm{I}^{MOT}_{Cool}(\rm{X}_3)$. In combination with Figs. \ref{fig3}\textbf{b} and \textbf{c}, it is evident that the parameter distributions for these two variables approach the upper limits of their respective ranges. The ranges are determined by the experimental hardware system. According to atomic cooling theory\cite{footAtomicPhysics2011}, the force experienced by an atom in the MOT can be given by: 

\begin{equation}
\label{FMOT}
F_{\mathrm{MOT}} = -4\hbar k^2 \frac{\mathrm{I}^{MOT}_{Cool}}{\mathrm{I}_{\mathrm{sat}}} \frac{-2\delta/\Gamma}{\left[1+\left(2\delta/\Gamma\right)^2\right]^2} v - 4\hbar k^2 \frac{\mathrm{I}^{MOT}_{Cool}}{\mathrm{I}_{\mathrm{sat}}} \frac{-2\delta/\Gamma}{\left[1+\left(2\delta/\Gamma\right)^2\right]^2}\frac{g\mu_\mathrm{B}}{k\hbar}\frac{\mathrm{d}B}{\mathrm{d}z}z,
\end{equation} 
where $\hbar k$ represents the photon momentum, $\Gamma$ is the natural linewidth, $I_{\mathrm{sat}}$ denotes the saturation absorption intensity, and $\delta$ is the detuning, representing the difference between the laser frequency and the atomic resonance frequency. For transitions from the hyperfine level $|F, M_F\rangle$ to $|F', M_{F'}\rangle$, $g = g_F' M_{F'} - g_F M_F$. The quantity $\mu_\mathrm{B}$ is the Bohr magneton, and $\mathrm{d}B/\mathrm{d}z$ is the magnetic field gradient, which increases proportionally with the applied voltage of the gradient magnetic field $\rm{X}_5$. The Zeeman effect induces an imbalance in the radiation force, creating a restoring force. Under typical operating conditions, the atom experiences overdamped simple harmonic motion. Atoms entering the overlapping region of the laser beams are decelerated, and the position-dependent force drives cold atoms toward the center of the trap. According to the Eq. \ref{FMOT}, it can be inferred that higher values of  $\mathrm{I}^{MOT}_{Cool}(\rm{X}_3)$ lead to stronger damping effects, enabling more efficient cooling of atoms. Additionally, A larger $\rm{X}_5$ generates a stronger magnetic field gradient, ultimately enhancing the trapping force. The stronger trapping force results in greater confinement of cold atoms at the MOT center. As illustrate in Fig. \ref{fig3}\textbf{a}, the correlation analysis results align well with established theoretical models.

Similarly, we perform a correlation analysis based on PGC experiment results to investigate the factors influencing both the temperature and atom number of the cold atomic ensemble. Fig. \ref{fig3}\textbf{d} presents the correlation matrix, which quantifies the relationships between various experimental parameters. Our focus is on identifying key parameters impacting temperature and atom number, as well as understanding the interplay between these two objectives. Specifically, the final atomic temperature is strongly affected by the frequency of high-detuned PGC and the endpoint of cooling light intensity attenuation. In addition, the atom number is influenced by the re-lighting time interval between MOT and PGC, the initial power of the reactivated cooling light, and the number of steps in the power attenuation process. Furthermore, according to statistical mechanics principles, the temperature of the atomic ensemble depends on the velocity distribution of the atoms. A lower atom number often indicates that high-speed atoms have escaped, leaving only low-speed atoms in the ensemble, which explains the observed inverse relationship between temperature and atom number. Figs. \ref{fig3}\textbf{e-f} display parameter distributions along with probability density plots. The purple histograms and curves correspond to the full dataset of 500 experiments, while the red histograms and curves represent data points with temperatures below $10~\rm{\mu K}$. Notably, the trends for both datasets are largely consistent, demonstrating that the Gaussian process regression model developed through MBO effectively characterizes the experimental system and accurately identifies parameter distributions meeting operational requirements. These findings enable further refinement of the optimization range, thereby enhancing the efficiency of parameter scanning.

\subsection{Autonomous problem diagnosis}

\begin{figure}[h]
\centering
\includegraphics[width=0.9\textwidth]{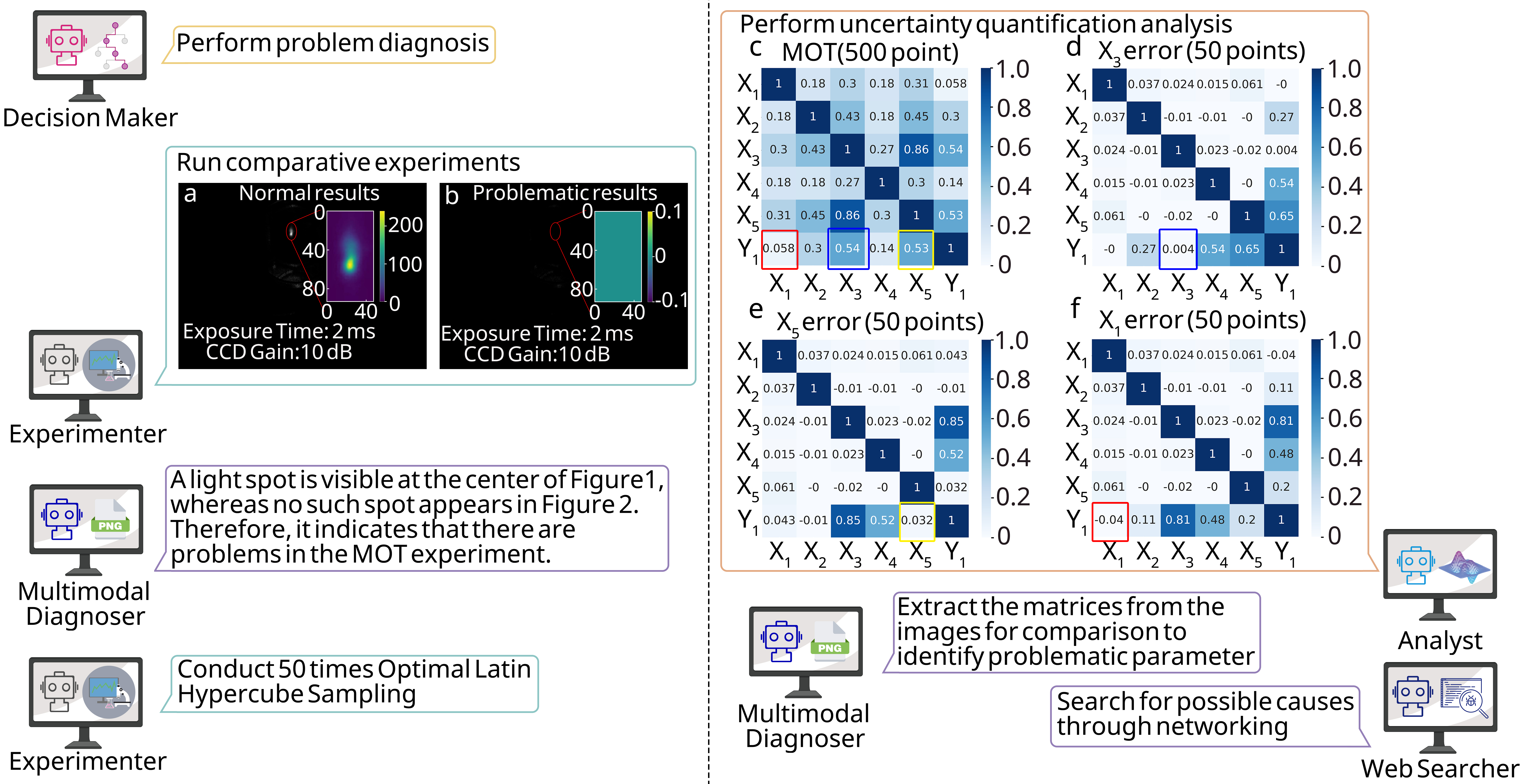}
\caption{\textbf{Fault detection and diagnosis in the experimental system} $\quad$When operational anomalies occur, the Decision Maker initiates comparative experiments, and the Multimodal Diagnoser contrasts normal outcomes (\textbf{a}) with anomalous results (\textbf{b}) to pinpoint the faulty sub-experiment. For the identified sub-experiment, 50 iterations of optimal Latin hypercube sampling are executed, followed by uncertainty quantification analysis of the resulting data. \textbf{c}, Baseline correlation matrix derived from 500 Bayesian optimization runs during MOT optimization. \textbf{d}-\textbf{f}, Correlation matrices under specific fault conditions: cooling light intensity malfunction (\textbf{d}), gradient magnetic field voltage anomaly (\textbf{e}), and optical phase-locked loop reference frequency deviation (\textbf{f}). Parameters in \textbf{c}-\textbf{f} correspond to those defined in Fig. \ref{fig3}\textbf{a}. Through the Multimodal Diagnoser, the problematic parameter is precisely identified; the online retriever is then activated to analyze root causes, enabling efficient fault resolution. 
}\label{fig4}
\end{figure}

After the optimization experiment is completed, the system transitions to the self-sustained running stage. During this stage, all sub-experiments are executed sequentially according to the list order, utilizing the optimal parameters derived from the optimization process for parameter fixation and result preservation. If significant deviations in experimental results are detected, the framework enters the problem diagnosis stage. In this stage, sub-experiments are rerun with the same parameters, and their results, together with those from the self-sustained running stage, are submitted to a multimodal diagnostic system for analysis. Fig. \ref{fig4}\textbf{a} displays the MOT experimental results during the self-sustained running stage, whereas Fig. \ref{fig4}\textbf{b} shows the MOT experimental results during the problem diagnosis stage. The Multimodal Diagnoser, which possesses image analysis capabilities, identifies differences by comparing the two sets of images, thereby localizing the problematic sub-experiment.

After isolating the problematic sub-experiment, the Decision Maker instructs the Experimenter to conduct 50 times optimal Latin hypercube sampling. The resulting experimental data are subjected to correlation analysis. The Multimodal Diagnoser extracts correlation matrix values from the correlation analysis result graph and compares them with those obtained during the optimization stage to identify potentially problematic parameters. Fig. \ref{fig4}\textbf{c} displays the correlation analysis results from the optimization stage, while the remaining figures depict the results after artificial errors were introduced. The blue box highlights the correlation coefficient between $\rm{X_3}$ and $\rm{Y_1}$, the yellow box highlights that between $\rm{X_5}$ and $\rm{Y_1}$, and the red box highlights that between $\rm{X_1}$ and $\rm{Y_1}$. In the first experiment, we fix the cooling light intensity ($\rm{X_3}$) at 4 mW/$\rm{cm^2}$. Fig. \ref{fig4}\textbf{d} clearly shows that the originally strong correlation for cooling light intensity decreases by two orders of magnitude, while the other parameters exhibit little change. In the second set of experiments, we reduce the actual magnetic field voltage ($\rm{X_5}$) to half its set value, drastically diminishing the trapping force. As shown in Fig. \ref{fig4}\textbf{e}, the correlations of most variables remain at the same order of magnitude, whereas the correlations for voltage-controlled oscillator frequency difference and gradient magnetic field voltage decrease by more than an order of magnitude, thereby reducing the list of suspect parameters from five to two.  In the third set of experiments, we fix the optical phase-locked loop reference frequency ($\rm{X_1}$) to 38.6 MHz. Since this parameter is inherently weakly correlated, Fig. \ref{fig4}\textbf{f} confirms its minimal impact on the results. By leveraging the Multimodal Diagnoser, the problematic parameter is pinpointed, and the online retriever is invoked to analyze the root cause of the fault, enabling effective problem detection and diagnosis.

\section{Discussion}\label{sec12}
To address the challenges of high construction costs and the difficulties associated with fault detection and diagnosis in modern quantum sensors, we propose a multi-disciplinary, coupled dynamic tuning framework built on a multi-agent system that leverages large language models. This framework is specifically designed for applications in the quantum domain.
%In contrast to prior methods that primarily focused on experimental optimization,
The proposed framework facilitates highly efficient autonomous operation across the entire workflow, including experimental optimization, problem detection, and diagnostic processes. By conducting only a few hundred experimental trials, the framework can identify solutions for single-objective optimization or determine the Pareto frontier for multi-objective optimization. Furthermore, after completing uncertainty quantification modeling, only 50 optimal Latin hypercube samples are required to accurately pinpoint critical problem parameters. To address the limitations of existing methods in knowledge accumulation, the framework incorporates a local knowledge base and supports network-based retrieval, enabling effective alignment and incremental knowledge acquisition.

However, the framework currently depends on online API calls to access large language models, which restrict its offline applications. Fortunately, advances in localized inference models (e.g., qwq-32B) on consumer-grade hardware, such as a single 24GB GPU, are gradually bridging the performance gap with commercial large-scale models. By integrating further optimizations in prompt engineering, the framework holds the potential for local deployment on standard consumer laptops. This development could ultimately enable the autonomous operation of quantum sensors in field applications.

\bmhead{Supplementary information}

%If your article has accompanying supplementary file/s please state so here. 

%Authors reporting data from electrophoretic gels and blots should supply the full unprocessed scans for key as part of their Supplementary information. This may be requested by the editorial team/s if it is missing.

%Please refer to Journal-level guidance for any specific requirements.

\bmhead{Acknowledgements}
This work was supported by National Natural Science Foundation of China (Grant No. 12404556, 62103426) and Hunan provincial major sci-tech program (2023ZJ1010) and.

\bibliography{sn-bibliography}% common bib file

\begin{thebibliography}{10}
\expandafter\ifx\csname url\endcsname\relax
  \def\url#1{\burl{#1}}\fi
\expandafter\ifx\csname urlprefix\endcsname\relax\def\urlprefix{URL }\fi
\providecommand{\bibinfo}[2]{#2}
\providecommand{\eprint}[2][]{\url{#2}}
\providecommand{\doi}[1]{\url{https://doi.org/#1}}
\bibcommenthead

\bibitem{wiemanAtomCoolingTrapping1999}
\bibinfo{author}{Wieman, C.~E.}
\newblock \bibinfo{title}{Atom cooling, trapping, and quantum manipulation}.
\newblock \emph{\bibinfo{journal}{Rev. Mod. Phys.}} \textbf{\bibinfo{volume}{71}}, \bibinfo{pages}{S253--S262} (\bibinfo{year}{1999}).

\bibitem{beckerSpaceborneBoseEinstein2018}
\bibinfo{author}{Becker, D.} \emph{et~al.}
\newblock \bibinfo{title}{Space-borne {{Bose}}--{{Einstein}} condensation for precision interferometry}.
\newblock \emph{\bibinfo{journal}{Nature}} \textbf{\bibinfo{volume}{562}}, \bibinfo{pages}{391--395} (\bibinfo{year}{2018}).

\bibitem{schreckLaserCoolingQuantum2021}
\bibinfo{author}{Schreck, F.} \& \bibinfo{author}{Druten, K.~V.}
\newblock \bibinfo{title}{Laser cooling for quantum gases}.
\newblock \emph{\bibinfo{journal}{Nat. Phys.}} \textbf{\bibinfo{volume}{17}}, \bibinfo{pages}{1296--1304} (\bibinfo{year}{2021}).

\bibitem{Yuan_2023}
\bibinfo{author}{Yuan, J.} \emph{et~al.}
\newblock \bibinfo{title}{Quantum sensing of microwave electric fields based on rydberg atoms}.
\newblock \emph{\bibinfo{journal}{Rep. Prog. Phys.}} \textbf{\bibinfo{volume}{86}}, \bibinfo{pages}{106001} (\bibinfo{year}{2023}).

\bibitem{huangEntanglementenhancedQuantumMetrology2024}
\bibinfo{author}{Huang, J.}, \bibinfo{author}{Zhuang, M.} \& \bibinfo{author}{Lee, C.}
\newblock \bibinfo{title}{Entanglement-enhanced quantum metrology: {{From}} standard quantum limit to {{Heisenberg}} limit}.
\newblock \emph{\bibinfo{journal}{Appl. Phys. Rev.}} \textbf{\bibinfo{volume}{11}}, \bibinfo{pages}{031302} (\bibinfo{year}{2024}).

\bibitem{SalducciSA2024}
\bibinfo{author}{Salducci, C.} \emph{et~al.}
\newblock \bibinfo{title}{Quantum sensing of acceleration and rotation by interfering magnetically launched atoms}.
\newblock \emph{\bibinfo{journal}{Sci. Adv.}} \textbf{\bibinfo{volume}{10}}, \bibinfo{pages}{eadq4498} (\bibinfo{year}{2024}).

\bibitem{pandaMeasuringGravitationalAttraction2024}
\bibinfo{author}{Panda, C.~D.} \emph{et~al.}
\newblock \bibinfo{title}{Measuring gravitational attraction with a lattice atom interferometer}.
\newblock \emph{\bibinfo{journal}{Nature}} \textbf{\bibinfo{volume}{631}}, \bibinfo{pages}{515--520} (\bibinfo{year}{2024}).

\bibitem{JunPRL2024}
\bibinfo{author}{Ye, J.} \& \bibinfo{author}{Zoller, P.}
\newblock \bibinfo{title}{Essay: Quantum sensing with atomic, molecular, and optical platforms for fundamental physics}.
\newblock \emph{\bibinfo{journal}{Phys. Rev. Lett.}} \textbf{\bibinfo{volume}{132}}, \bibinfo{pages}{190001} (\bibinfo{year}{2024}).

\bibitem{barrettDualMatterwaveInertial2016}
\bibinfo{author}{Barrett, B.} \emph{et~al.}
\newblock \bibinfo{title}{Dual matter-wave inertial sensors in weightlessness}.
\newblock \emph{\bibinfo{journal}{Nat. Commun.}} \textbf{\bibinfo{volume}{7}} (\bibinfo{year}{2016}).

\bibitem{bidelAbsoluteMarineGravimetry2018}
\bibinfo{author}{Bidel, Y.} \emph{et~al.}
\newblock \bibinfo{title}{Absolute marine gravimetry with matter-wave interferometry}.
\newblock \emph{\bibinfo{journal}{Nat. Commun.}} \textbf{\bibinfo{volume}{9}}, \bibinfo{pages}{627} (\bibinfo{year}{2018}).

\bibitem{Bongs:2019exm}
\bibinfo{author}{Bongs, K.} \emph{et~al.}
\newblock \bibinfo{title}{{Taking atom interferometric quantum sensors from the laboratory to real-world applications}}.
\newblock \emph{\bibinfo{journal}{Nat. Rev. Phys.}} \textbf{\bibinfo{volume}{1}}, \bibinfo{pages}{731--739} (\bibinfo{year}{2019}).

\bibitem{XuejianSA2019}
\bibinfo{author}{Wu, X.} \emph{et~al.}
\newblock \bibinfo{title}{Gravity surveys using a mobile atom interferometer}.
\newblock \emph{\bibinfo{journal}{Sci. Adv.}} \textbf{\bibinfo{volume}{5}}, \bibinfo{pages}{eaax0800} (\bibinfo{year}{2019}).

\bibitem{strayQuantumSensingGravity2022}
\bibinfo{author}{Stray, B.} \emph{et~al.}
\newblock \bibinfo{title}{Quantum sensing for gravity cartography}.
\newblock \emph{\bibinfo{journal}{Nature}} \textbf{\bibinfo{volume}{602}}, \bibinfo{pages}{590--594} (\bibinfo{year}{2022}).

\bibitem{darmagnacdecastanetAtomInterferometryArbitrary2024}
\bibinfo{author}{{d'Armagnac De Castanet}, Q.} \emph{et~al.}
\newblock \bibinfo{title}{Atom interferometry at arbitrary orientations and rotation rates}.
\newblock \emph{\bibinfo{journal}{Nat. Commun.}} \textbf{\bibinfo{volume}{15}}, \bibinfo{pages}{6406} (\bibinfo{year}{2024}).

\bibitem{dengColdAtomMicrowave2024}
\bibinfo{author}{Deng, S.} \emph{et~al.}
\newblock \bibinfo{title}{Cold atom microwave clock based on intracavity cooling in {{China}} space station}.
\newblock \emph{\bibinfo{journal}{npj Microgravity}} \textbf{\bibinfo{volume}{10}}, \bibinfo{pages}{66} (\bibinfo{year}{2024}).

\bibitem{zhaiAirborneAbsoluteGravity2025}
\bibinfo{author}{Zhai, C.} \emph{et~al.}
\newblock \bibinfo{title}{Airborne absolute gravity measurements based on quantum gravimeter}.
\newblock \emph{\bibinfo{journal}{Acta Phys. Sin.}} \textbf{\bibinfo{volume}{74}}, \bibinfo{pages}{070302} (\bibinfo{year}{2025}).

\bibitem{26vs-4t7p2025}
\bibinfo{author}{Li, C.-Y.} \emph{et~al.}
\newblock \bibinfo{title}{Drift-free continuous gravity measurement and application analysis of a high-precision atom gravimeter}.
\newblock \emph{\bibinfo{journal}{Phys. Rev. Appl.}} \textbf{\bibinfo{volume}{24}}, \bibinfo{pages}{014045} (\bibinfo{year}{2025}).

\bibitem{LiNSR2025}
\bibinfo{author}{Li, J.} \emph{et~al.}
\newblock \bibinfo{title}{Realization of a cold atom gyroscope in space}.
\newblock \emph{\bibinfo{journal}{Natl. Sci. Rev.}} \textbf{\bibinfo{volume}{12}}, \bibinfo{pages}{nwaf012} (\bibinfo{year}{2025}).

\bibitem{DegenRevModPhys2017}
\bibinfo{author}{Degen, C.~L.}, \bibinfo{author}{Reinhard, F.} \& \bibinfo{author}{Cappellaro, P.}
\newblock \bibinfo{title}{Quantum sensing}.
\newblock \emph{\bibinfo{journal}{Rev. Mod. Phys.}} \textbf{\bibinfo{volume}{89}}, \bibinfo{pages}{035002} (\bibinfo{year}{2017}).

\bibitem{geraciSensitivityAtomInterferometry2016}
\bibinfo{author}{Geraci, A.~A.} \& \bibinfo{author}{Derevianko, A.}
\newblock \bibinfo{title}{Sensitivity of atom interferometry to ultralight scalar field dark matter}.
\newblock \emph{\bibinfo{journal}{Phys. Rev. Lett.}} \textbf{\bibinfo{volume}{117}}, \bibinfo{pages}{261301} (\bibinfo{year}{2016}).

\bibitem{saywellEnhancingSensitivityAtominterferometric2023a}
\bibinfo{author}{Saywell, J.~C.} \emph{et~al.}
\newblock \bibinfo{title}{Enhancing the sensitivity of atom-interferometric inertial sensors using robust control}.
\newblock \emph{\bibinfo{journal}{Nat. Commun.}} \textbf{\bibinfo{volume}{14}}, \bibinfo{pages}{7626} (\bibinfo{year}{2023}).

\bibitem{2016NatSR...625890W}
\bibinfo{author}{{Wigley}, P.~B.} \emph{et~al.}
\newblock \bibinfo{title}{{Fast machine-learning online optimization of ultra-cold-atom experiments}}.
\newblock \emph{\bibinfo{journal}{Sci. Rep.}} \textbf{\bibinfo{volume}{6}}, \bibinfo{pages}{25890} (\bibinfo{year}{2016}).

\bibitem{nakamuraNonstandardTrajectoriesFound2019}
\bibinfo{author}{Nakamura, I.}, \bibinfo{author}{Kanemura, A.}, \bibinfo{author}{Nakaso, T.}, \bibinfo{author}{Yamamoto, R.} \& \bibinfo{author}{Fukuhara, T.}
\newblock \bibinfo{title}{Non-standard trajectories found by machine learning for evaporative cooling of {\textsuperscript{87}} {{Rb}} atoms}.
\newblock \emph{\bibinfo{journal}{Opt. Express}} \textbf{\bibinfo{volume}{27}}, \bibinfo{pages}{20435} (\bibinfo{year}{2019}).

\bibitem{Barker_2020}
\bibinfo{author}{Barker, A.~J.} \emph{et~al.}
\newblock \bibinfo{title}{Applying machine learning optimization methods to the production of a quantum gas}.
\newblock \emph{\bibinfo{journal}{Mach. Learn.: Sci. Technol.}} \textbf{\bibinfo{volume}{1}}, \bibinfo{pages}{015007} (\bibinfo{year}{2020}).

\bibitem{maAcceleratedBayesianOptimization2025}
\bibinfo{author}{Ma, X.} \emph{et~al.}
\newblock \bibinfo{title}{Accelerated bayesian optimization in deep cooling atoms} (\bibinfo{year}{2025}).
\newblock \eprint{arXiv:2412.11793}.

\bibitem{weidnerAtomInterferometryUsing2017}
\bibinfo{author}{Weidner, C.~A.}, \bibinfo{author}{Yu, H.}, \bibinfo{author}{Kosloff, R.} \& \bibinfo{author}{Anderson, D.~Z.}
\newblock \bibinfo{title}{Atom interferometry using a shaken optical lattice}.
\newblock \emph{\bibinfo{journal}{Phys. Rev. A}} \textbf{\bibinfo{volume}{95}}, \bibinfo{pages}{043624} (\bibinfo{year}{2017}).

\bibitem{2018NatCo...9.4360T}
\bibinfo{author}{{Tranter}, A.~D.} \emph{et~al.}
\newblock \bibinfo{title}{{Multiparameter optimisation of a magneto-optical trap using deep learning}}.
\newblock \emph{\bibinfo{journal}{Nat. Commun.}} \textbf{\bibinfo{volume}{9}}, \bibinfo{pages}{4360} (\bibinfo{year}{2018}).

\bibitem{AlexeyPNAS2018}
\bibinfo{author}{Melnikov, A.~A.} \emph{et~al.}
\newblock \bibinfo{title}{Active learning machine learns to create new quantum experiments}.
\newblock \emph{\bibinfo{journal}{Proc. Natl. Acad. Sci. U.S.A.}} \textbf{\bibinfo{volume}{115}}, \bibinfo{pages}{1221--1226} (\bibinfo{year}{2018}).

\bibitem{2023AdPho...5a6005C}
\bibinfo{author}{{Cimini}, V.} \emph{et~al.}
\newblock \bibinfo{title}{{Deep reinforcement learning for quantum multiparameter estimation}}.
\newblock \emph{\bibinfo{journal}{Adv. Photonics}} \textbf{\bibinfo{volume}{5}}, \bibinfo{pages}{016005} (\bibinfo{year}{2023}).

\bibitem{chihReinforcementLearningRotation2024}
\bibinfo{author}{Chih, L.-Y.} \& \bibinfo{author}{Holland, M.}
\newblock \bibinfo{title}{Reinforcement learning for rotation sensing with ultracold atoms in an optical lattice}.
\newblock \emph{\bibinfo{journal}{Phys. Rev. Res.}} \textbf{\bibinfo{volume}{6}}, \bibinfo{pages}{043191} (\bibinfo{year}{2024}).

\bibitem{2024NatCo..15.8532R}
\bibinfo{author}{{Reinschmidt}, M.}, \bibinfo{author}{{Fort{\'a}gh}, J.}, \bibinfo{author}{{G{\"u}nther}, A.} \& \bibinfo{author}{{Volchkov}, V.~V.}
\newblock \bibinfo{title}{{Reinforcement learning in cold atom experiments}}.
\newblock \emph{\bibinfo{journal}{Nat. Commun.}} \textbf{\bibinfo{volume}{15}}, \bibinfo{pages}{8532} (\bibinfo{year}{2024}).

\bibitem{Liang24}
\bibinfo{author}{Liang, C.} \emph{et~al.}
\newblock \bibinfo{title}{Multi-parameter optimization of polarization gradient cooling for 87rb atoms based on reinforcement learning}.
\newblock \emph{\bibinfo{journal}{Opt. Express}} \textbf{\bibinfo{volume}{32}}, \bibinfo{pages}{40364--40374} (\bibinfo{year}{2024}).

\bibitem{wangScientificDiscoveryAge2023}
\bibinfo{author}{Wang, H.} \emph{et~al.}
\newblock \bibinfo{title}{Scientific discovery in the age of artificial intelligence}.
\newblock \emph{\bibinfo{journal}{Nature}} \textbf{\bibinfo{volume}{620}}, \bibinfo{pages}{47--60} (\bibinfo{year}{2023}).

\bibitem{singh_progprompt_2023}
\bibinfo{author}{Singh, I.} \emph{et~al.}
\newblock \bibinfo{title}{{ProgPrompt}: program generation for situated robot task planning using large language models}.
\newblock \emph{\bibinfo{journal}{Auton. Robot}} \textbf{\bibinfo{volume}{47}}, \bibinfo{pages}{999--1012} (\bibinfo{year}{2023}).

\bibitem{boiko_autonomous_2023}
\bibinfo{author}{Boiko, D.~A.}, \bibinfo{author}{MacKnight, R.}, \bibinfo{author}{Kline, B.} \& \bibinfo{author}{Gomes, G.}
\newblock \bibinfo{title}{Autonomous chemical research with large language models}.
\newblock \emph{\bibinfo{journal}{Nature}} \textbf{\bibinfo{volume}{624}}, \bibinfo{pages}{570--578} (\bibinfo{year}{2023}).

\bibitem{gaoEmpoweringBiomedicalDiscovery2024}
\bibinfo{author}{Gao, S.} \emph{et~al.}
\newblock \bibinfo{title}{Empowering biomedical discovery with {{AI}} agents}.
\newblock \emph{\bibinfo{journal}{Cell}} \textbf{\bibinfo{volume}{187}}, \bibinfo{pages}{6125--6151} (\bibinfo{year}{2024}).

\bibitem{Dong2024}
\bibinfo{author}{Dong, Y.}, \bibinfo{author}{Jiang, X.}, \bibinfo{author}{Jin, Z.} \& \bibinfo{author}{Li, G.}
\newblock \bibinfo{title}{Self-collaboration code generation via chatgpt}.
\newblock \emph{\bibinfo{journal}{ACM Trans. Softw. Eng. Methodol.}} \textbf{\bibinfo{volume}{33}} (\bibinfo{year}{2024}).

\bibitem{Chen2024}
\bibinfo{author}{Chen, Y.}, \bibinfo{author}{Arkin, J.}, \bibinfo{author}{Zhang, Y.}, \bibinfo{author}{Roy, N.} \& \bibinfo{author}{Fan, C.}
\newblock \bibinfo{title}{Scalable multi-robot collaboration with large language models: Centralized or decentralized systems?} (\bibinfo{year}{2024}).
\newblock \bibinfo{note}{Paper presented at the 2024 IEEE International Conference on Robotics and Automation (ICRA), 4311-4317 2024}.

\bibitem{sandersBiologicalResearchSelfdriving2023}
\bibinfo{author}{Sanders, L.~M.} \emph{et~al.}
\newblock \bibinfo{title}{Biological research and self-driving labs in deep space supported by artificial intelligence}.
\newblock \emph{\bibinfo{journal}{Nat. Mach. Intell.}} \textbf{\bibinfo{volume}{5}}, \bibinfo{pages}{208--219} (\bibinfo{year}{2023}).

\bibitem{m_bran_augmenting_2024}
\bibinfo{author}{M.~Bran, A.} \emph{et~al.}
\newblock \bibinfo{title}{Augmenting large language models with chemistry tools}.
\newblock \emph{\bibinfo{journal}{Nat. Mach. Intell.}} \textbf{\bibinfo{volume}{6}}, \bibinfo{pages}{525--535} (\bibinfo{year}{2024}).

\bibitem{luoLargeLanguageModels2024}
\bibinfo{author}{Luo, X.} \emph{et~al.}
\newblock \bibinfo{title}{Large language models surpass human experts in predicting neuroscience results}.
\newblock \emph{\bibinfo{journal}{Nat. Hum. Behav}}  (\bibinfo{year}{2024}).

\bibitem{ruanAutomaticEndtoendChemical2024}
\bibinfo{author}{Ruan, Y.} \emph{et~al.}
\newblock \bibinfo{title}{An automatic end-to-end chemical synthesis development platform powered by large language models}.
\newblock \emph{\bibinfo{journal}{Nat. Commun.}} \textbf{\bibinfo{volume}{15}}, \bibinfo{pages}{10160} (\bibinfo{year}{2024}).

\bibitem{jablonkaLeveragingLargeLanguage2024}
\bibinfo{author}{Jablonka, K.~M.}, \bibinfo{author}{Schwaller, P.}, \bibinfo{author}{{Ortega-Guerrero}, A.} \& \bibinfo{author}{Smit, B.}
\newblock \bibinfo{title}{Leveraging large language models for predictive chemistry}.
\newblock \emph{\bibinfo{journal}{Nat. Mach. Intell.}} \textbf{\bibinfo{volume}{6}}, \bibinfo{pages}{161--169} (\bibinfo{year}{2024}).

\bibitem{wu_leveraging_2024}
\bibinfo{author}{Wu, Z.} \emph{et~al.}
\newblock \bibinfo{title}{Leveraging language model for advanced multiproperty molecular optimization via prompt engineering}.
\newblock \emph{\bibinfo{journal}{Nat. Mach. Intell.}} \bibinfo{pages}{1359--1369} (\bibinfo{year}{2024}).

\bibitem{tshitoyanUnsupervisedWordEmbeddings2019}
\bibinfo{author}{Tshitoyan, V.} \emph{et~al.}
\newblock \bibinfo{title}{Unsupervised word embeddings capture latent knowledge from materials science literature}.
\newblock \emph{\bibinfo{journal}{Nature}} \textbf{\bibinfo{volume}{571}}, \bibinfo{pages}{95--98} (\bibinfo{year}{2019}).

\bibitem{melkoLanguageModelsQuantum2024}
\bibinfo{author}{Melko, R.~G.} \& \bibinfo{author}{Carrasquilla, J.}
\newblock \bibinfo{title}{Language models for quantum simulation}.
\newblock \emph{\bibinfo{journal}{Nat. Comput. Sci.}} \textbf{\bibinfo{volume}{4}}, \bibinfo{pages}{11--18} (\bibinfo{year}{2024}).

\bibitem{ghafarollahiSciAgentsAutomatingScientific2024}
\bibinfo{author}{Ghafarollahi, A.} \& \bibinfo{author}{Buehler, M.~J.}
\newblock \bibinfo{title}{Sciagents: Automating scientific discovery through bioinspired multi-agent intelligent graph reasoning}.
\newblock \emph{\bibinfo{journal}{Adv. Mater.}} \bibinfo{pages}{2413523} (\bibinfo{year}{2024}).

\bibitem{footAtomicPhysics2011}
\bibinfo{author}{Foot, C.~J.}
\newblock \emph{\bibinfo{title}{Atomic Physics}} No.~\bibinfo{number}{7} in \bibinfo{series}{Oxford Master Series in Physics {{Atomic}}, Optical, and Laser Physics} (\bibinfo{publisher}{Oxford Univ. Press}, \bibinfo{address}{Oxford}, \bibinfo{year}{2011}).

\end{thebibliography}
%% if required, the content of .bbl file can be included here once bbl is generated
%%\input sn-article.bbl

\end{document}